\documentstyle[12pt,epsfig]{article}
\def\lbldef#1#2{\expandafter\gdef\csname #1\endcsname {#2}}

\def\href#1#2{#2}  

\begin{document}
\baselineskip=15.5pt
\pagestyle{plain}
\setcounter{page}{1}

\begin{titlepage}

\begin{flushright}
CERN-TH/99-364\\
hep-th/9911218
\end{flushright}
\vspace{10 mm}

\begin{center}
{\Large Solitons in Brane Worlds}

\vspace{5mm}

\end{center}

\vspace{5 mm}

\begin{center}
{\large Donam Youm\footnote{E-mail: Donam.Youm@cern.ch}}

\vspace{3mm}

Theory Division, CERN, CH-1211, Geneva 23, Switzerland

\end{center}

\vspace{1cm}

\begin{center}
{\large Abstract}
\end{center}

\noindent

We study some aspects of dilatonic domain walls in relation to the idea on 
the noncompact internal space.  We find that the warp factor in the spacetime 
metric increases as one moves away from the domain wall for all the 
supersymmetric dilatonic domain wall solutions obtained from the 
(intersecting) BPS branes in string theories through toroidal 
compactifications, unlike the case of the Randall-Sundrum model.  On 
the other hand, when the dilaton coupling parameter $a$ for the 
$D$-dimensional extreme dilatonic domain wall takes the values 
$a^2<4/(D-2)^2$, the Kaluza-Klein spectrum of graviton has the same structure 
as that of the Randall-Sundrum model (and the warp factor decreases in the 
finite interval around the dilatonic domain wall), thereby implying the 
possibility of extending the Randall-Sundrum model to the $a^2<4/(D-2)^2$ 
case.  We construct fully localized solutions describing extreme dilatonic 
branes within extreme dilatonic domain walls and the supersymmetric branes 
within the supersymmetric domain walls of string theories.  These solutions 
are valid in any region of spacetime, not just in the region close to the 
domain walls.  

\vspace{1cm}
\begin{flushleft}
CERN-TH/99-364\\
November, 1999
\end{flushleft}
\end{titlepage}
\newpage

\section{Introduction}

It was Kaluza \cite{kal} who showed that the Einstein's theory of 
four-dimensional gravity and the Maxwell's theory of electromagnetism 
can be unified within the five-dimensional theory of general relativity. 
In his derivation of his model, Kaluza avoided the question of unobservableness
of such extra fifth coordinate simply by assuming that all the fields do not 
depend on the extra coordinates.  Later, Klein \cite{kle} explained the 
unobservableness of and the independence of physical quantities on the extra 
coordinate by assuming the extra spatial dimension to be too small to be 
observed.  Namely, if one assumes the extra dimension to be a compact 
manifold, i.e., a circle in the Klein case, then all the fields can be 
Fourier-expanded and the Fourier coefficients are identified as fields in 
compactified lower dimensions.  Since mass scale of such Fourier modes, 
called the Kaluza-Klein (KK) modes, is inversely proportional to the size 
of compactification manifold, if the size of the manifold becomes very small, 
then the extra massive KK modes become too heavy to be observed.  So, in the 
limit of very small compactification manifold, one can effectively keep only 
zero modes of the KK spectrum, thereby leading to the original Kaluza's 
assumption that the physical quantities simply do not depend on the extra 
coordinate.  

This idea of Klein has been dominantly taken within particle physics, 
including the KK supergravity theories and superstring theories. 
However, the problem with the Klein's idea on small extra space is that 
it cannot explain why the extra space has to be markedly different in 
topology and in size.  One of alternative approaches which attempt to 
address this problem assumes that the extra space is on the same footing 
as our four-dimensional spacetime, namely the fields depend on the extra 
space which is now assumed to have infinite extend just like our 
four-dimensional spacetime or to be not very small.  However, the challenge 
of this approach is to explain why the extra space has not been observed.  
To avoid this difficulty, it was argued that particles are trapped inside of 
the four-dimensional hypersurface by a potential well 
\cite{jos,aka,rs1,rs2,vis,sq,lm,go1,go2,go3} or for topological reason 
\cite{gw}.  This idea was also proposed in an attempt to explain smallness 
of cosmological constant in our four-dimensional universe \cite{rs2,vis,sq} 
and to bring chiral fermions \cite{ch1,ch2,ch3,ch4,ch5,ch6,ch7} into the 
theory, which is not achievable from the conventional KK theory with compact 
internal space.  

This old idea on noncompact extra space recently has received 
revived attention, after it was found out \cite{ran1,ran2,ran3} by 
Randall and Sundrum (RS) that such approach can also be applied to solve 
the hierarchy problem of particle physics.  In the RS model, our 
four-dimensional world is regarded as a non-dilatonic 3-brane embedded in 
five-dimensional spacetime.  In the RS model, the potential due to gravity 
is repulsive, unlike the case of the previous works on trapping 
of matter inside of the four-dimensional hypersurface as discussed in 
the previous paragraph, because of the (exponentially) decreasing (rather 
than increasing) warp factor in the spacetime metric.   This exponentially 
decreasing warp factor makes the trapping of the gravity inside of the 
four-dimensional hypersurface possible.  In fact, it is also shown that 
such model with noncompact extra space reproduces \cite{go1,ran2} Newton's 
$1/r^2$ law of four-dimensional gravity with experimentally unobservable 
correction from the (extremely suppressed) contribution of the continuum 
of massive KK modes \cite{ran2}. 

The Ansatz for the five-dimensional spacetime metric used in the RS model is 
the one with the warp factor, where the four-dimensional part has the 
conformal factor depending on the extra spatial coordinate.  So, such metric 
can also be regarded as the metric of five-dimensional non-dilatonic domain 
wall, which is supported by the cosmological constant in five-dimensions.  
In this paper, we consider the case of dilatonic domain walls, where the 
cosmological constant term in the action is multiplied by the dilaton factor, 
with an arbitrary dilaton coupling parameter $a$ in any spacetime dimensions 
$D$.  (Some aspects of dilatonic generalization of the RS model are also 
studied, for example, in Refs. \cite{keh,hal,gcs}.)  We study trapping of 
matter near the domain wall and the KK spectrum of the graviton in the 
dilatonic domain wall background.  We find that for the supersymmetric 
dilatonic domain walls that can be obtained by toroidally compactifying the 
(intersecting) BPS branes in string theories, the warp factor in the metric 
increases as one moves away from the domain wall.  So, the RS-type scenario 
cannot be realized within such dilatonic domain walls.  This is along the 
same line as the previous works on supergravity embeddings 
\cite{beh,tow,gib,kall} of the RS model.  Namely, the BPS domain 
wall solutions in five-dimensional supergravity theories that have been 
constructed so far have exponentially increasing warp factor (as oppose to 
the exponentially decreasing warp factor of the RS model)
\footnote{I would like to thank Prof. Kallosh for bringing this 
point to my attention and giving suggestions for better presentation of the 
results in this paper, after the first version of the paper appeared on the 
preprint archive.}.  
This is shown explicitly first time in Ref. \cite{kall} within domain wall 
solutions in five-dimensional gauged $N=2$ supergravity.  On the other 
hand, we find that when the dilaton coupling parameter takes the values 
$a^2<4(D-1)/(D-2)^2$ the warp factor decreases in finite interval around 
the domain wall.  Particularly when $a^2<4/(D-2)^2$, the potential term in 
the Schr\"odinger equation describing the small fluctuation of the spacetime 
metric around the Minkowski sub-spacetime takes the qualitatively same form 
as that of the RS model, thereby implying the possibility of the extension 
of the RS model to the $a^2<4/(D-2)^2$ case.  Note, this cannot be obtained 
from the (intersecting) BPS branes in string theories through toroidal 
compactifications. We also construct fully localized solutions describing 
various extreme solitons living inside of the dilatonic domain walls, by 
applying the formalism studied in Ref. \cite{youm}.  These solutions are 
valid for any region in spacetime, not just in the region close to the 
domain walls.

The paper is organized as follows.  In section 2, we discuss aspects 
of dilatonic domain wall solution, especially in relation to the 
idea on noncompact internal space.  In section 3, we construct fully 
localized solutions describing extreme dilatonic branes within the extreme 
dilatonic domain walls.  In section 4, we construct fully localized 
supergravity solutions describing the BPS branes within the BPS 
domain walls in string theories.

\section{Dilatonic Domain Walls}

In this section, we discuss solution for a general dilatonic 
domain wall in $D$-dimensional spacetime and its role as an 
alternative to the conventional KK theory with noncompact internal 
space.   

We start by summarizing a general dilatonic $p$-brane solution in $D$ 
spacetime dimensions, since we are interested in obtaining solutions 
describing branes inside of the worldvolume of domain walls.  The 
Einstein-frame action for the dilatonic $p$-brane with the dilaton coupling 
parameter $a_p$ in $D$ spacetime dimensions is
\begin{equation}
S^E_p={1\over{2\kappa^2_D}}\int d^Dx\sqrt{-g^E}\left[{\cal R}_{g^E}
-{4\over{D-2}}(\partial\phi)^2-{1\over{2\cdot(p+2)!}}e^{2a_p\phi}F^2_{p+2}
\right],
\label{einpdrnact}
\end{equation}
where $\kappa_D$ is the $D$-dimensional Einstein gravitational constant, 
$\phi$ is the dilaton, $F_{p+2}=dA_{p+1}$ is the field strength of 
the $(p+1)$-form potential $A_{p+1}$.  The solution to the 
field equations of the Einstein-frame action (\ref{einpdrnact}) 
for the extreme dilatonic $p$-brane with the longitudinal coordinates 
${\bf x}=(x_1,...,x_p)$ and the transverse coordinates ${\bf y}=
(y_1,...,y_{D-p-1})$ located at ${\bf y}={\bf 0}$ has the following form:
\begin{eqnarray}
ds^2_E&=&H^{-{{4(D-p-3)}\over{(D-2)\Delta_p}}}_p\left[-dt^2+dx^2_1+
\cdots+dx^2_p\right]+H^{{4(p+1)}\over{(D-2)\Delta_p}}_p\left[dy^2_1+
\cdots+dy^2_{D-p-1}\right],
\cr
e^{2\phi}&=&H^{{(D-2)a_p}\over{\Delta_p}}_p,\ \ \ \ \ \ \ 
A_{tx_1...x_p}=1-H^{-1}_p,
\label{einpbrnsol}
\end{eqnarray}
where 
\begin{equation}
H_p=1+{{Q_p}\over{|{\bf y}|^{D-p-3}}},\ \ \ \ \ \ 
\Delta_p={{(D-2)a^2_p}\over 2}+{{2(p+1)(D-p-3)}\over{D-2}}.
\label{pbrndefs}
\end{equation}

For general values of the dilaton coupling parameter $a_p$, the 
extreme solution (\ref{einpbrnsol}) is not supersymmetric.  Only the 
extreme solutions with the specific values of the dilaton coupling 
parameter $a_p$ are supersymmetric.  For the consistently truncated 
supergravity action of the form (\ref{einpdrnact}) obtained by compactifying 
the eleven-dimensional supergravity on $(S^1)^{11-D}$ has the dilaton coupling 
$a_p$ such that $\Delta_p=4/N$ with a positive integer $N$ \cite{lp1,lp2}.  
Here, the field strength $F_{p+2}$ is a linear combination of $N$ original 
$(p+2)$-form field strengths with the same Page charges and the dilaton 
$\phi$ is a linear-combination of the original (dilatonic) scalars.
The extreme dilatonic $p$-brane solution (\ref{einpbrnsol}) with $\Delta_p=
4/N$ preserves at least $2^{-N}$ of the maximal symmetry.  In other words, 
all the dilatonic $p$-brane solutions obtained by toroidally compactifying 
intersecting $N$ numbers of BPS branes in string theories with equal charges 
have the form (\ref{einpbrnsol}) with $\Delta_p=4/N$.  Note, in this case, 
the ``dilaton'' $\phi$ in Eqs. (\ref{einpdrnact}) and (\ref{einpbrnsol}) is 
rather a linear combination of the dilaton and other scalars in string theory. 

The dilatonic domain wall with the dilaton coupling parameter $a$ in $D$ 
spacetime dimensions corresponds to the $p=D-2$ case of the general dilatonic 
$p$-brane discussed in the above.  Namely, a domain wall in $D$ spacetime 
dimensions can be regarded as a $(D-2)$-brane, electrically charged under 
the $(D-1)$-form potential $A_{D-1}$.  By applying a Poincar\'e dualization, 
one can replace the $D$-form field strength $F_D=dA_{D-1}$ by the 
cosmological constant $\Lambda$.  So, the action (\ref{einpdrnact}) with 
$p=D-2$ can be rewritten as
\begin{equation}
S^E_{\rm DW}={1\over{2\kappa^2_D}}\int d^Dx\sqrt{-g^E}\left[{\cal R}_{g^E}
-{4\over{D-2}}(\partial\phi)^2+e^{-2a\phi}\Lambda\right].
\label{cosdomanact}
\end{equation}
The Einstein-frame solution to the field equations of the action 
(\ref{cosdomanact}) for the extreme dilatonic domain wall located 
at $y=0$ has the following form:
\begin{eqnarray}
ds^2_E&=&H^{4\over{(D-2)\Delta}}\left[-dt^2+dx^2_1+\cdots+dx^2_{D-2}
\right]+H^{{4(D-1)}\over{(D-2)\Delta}}dy^2,
\cr
e^{2\phi}&=&H^{{(D-2)a}\over{\Delta}},\ \ \ \ \ \ 
A_{tx_1...x_{D-2}}=1-H^{-1},
\label{dmnwllsol}
\end{eqnarray}
where
\begin{equation}
H=1+ Q|y|,\ \ \ \ \ \ \ 
\Delta={{(D-2)a^2}\over{2}}-{{2(D-1)}\over{D-2}}.
\label{dmnwlldefs}
\end{equation}
Here, the parameter $Q$ in the harmonic function $H$ is related to the 
cosmological constant $\Lambda$ in the action (\ref{cosdomanact}) through
\begin{equation}
\Lambda=-{{2Q^2}\over{\Delta}}.
\label{coschrg}
\end{equation}

As mentioned in the previous paragraph, the supersymmetric
\footnote{Strictly speaking, supersymmetry nature of domain wall solutions 
cannot be fully answered until one considers full supergravity action with 
non-constant dilaton field (which makes the wall) and the explicit form 
of superpotential.} 
dilatonic domain wall solution (\ref{dmnwllsol}) that can be obtained by 
compactifying (intersecting) BPS branes in string theories on a torus has 
$\Delta=4/N$.  On the other hand, one can also obtain supersymmetric 
dilatonic domain wall solutions through different compactification 
procedure.  For example,  by compactifying the supersymmetric 
$D^{\prime}$-dimensional dilatonic $p$-brane action (\ref{einpdrnact}) with 
$\Delta_p=4/N$ on $S^{D^{\prime}-p-2}$ \cite{town,bbh}, one has the action 
(\ref{cosdomanact}) with $a$ and $\Delta$ given by:
\begin{eqnarray}
|a|&=&{2\over p}\sqrt{{2(D^{\prime}-2)-N(p+1)(D^{\prime}-p-3)}
\over{2(D^{\prime}-p-2)-N(D^{\prime}-p-3)}},
\cr
\Delta&=&-{{4(D^{\prime}-p-3)}\over{2(D^{\prime}-p-2)
-N(D^{\prime}-p-3)}}.
\label{adelexp}
\end{eqnarray}
Here, $D$ in Eqs. (\ref{cosdomanact}) $-$ (\ref{dmnwlldefs}) is given 
by $D=p+2$ for this type of compactification.  

Note, unlike the case of other type of branes, one can always remove the 
constant term in the above harmonic function $H$ for the domain wall by 
applying the coordinate translation along the $y$-direction, and the 
solution can be written with different powers of harmonic function 
through the coordinate transformations \cite{bs}.  In this paper we choose 
the transverse coordinate such that the solution has the standard form that 
would be obtained by taking the $p=D-2$ limit of dilatonic $p$-brane solutions.

In general, matter in the $D$-dimensional spacetime with the following 
form of metric with the warp factor $W(z)$:
\begin{equation}
\hat{g}_{\hat{\mu}\hat{\nu}}dx^{\hat{\mu}}dx^{\hat{\nu}}=W(z)
\left[-dt^2+dx^2_1+\cdots+dx^2_{D-2}\right]+dz^2
\label{warpmet}
\end{equation}
is trapped inside of the $(D-1)$-dimensional sub-spacetime with the 
coordinates $(x^{\mu})=(t,x_1,...,x_{D-2})$ for a suitable form of the 
warp factor $W(z)$.  To see \cite{vis} this classically, we consider a 
particle with the $D$-momentum $P=(P^0,P^i,P^z)$ moving in this spacetime.  
The Killing vectors for this spacetime are $\partial/\partial t$ and 
$\partial/\partial x^i$ ($i=1,...,D-2$).  So, the constants of motion for 
the particle are
\begin{eqnarray}
E&=&-\left(P,{\partial\over{\partial t}}\right)=-P^{\hat{\mu}}
\hat{g}_{\hat{\mu}\hat{\nu}}\left({\partial\over{\partial t}}
\right)^{\hat{\nu}}=P^0W(z),
\cr
p^i&=&\left(P,{\partial\over{\partial x^i}}\right)=P^{\hat{\mu}}
\hat{g}_{\hat{\mu}\hat{\nu}}\left({\partial\over{\partial x^i}}
\right)^{\hat{\nu}}=P^iW(z).
\label{cnstmtn}
\end{eqnarray}
The invariant mass $M$ of the particle in this spacetime is defined as
\begin{equation}
-M^2=P^{\hat{\mu}}\hat{g}_{\hat{\mu}\hat{\nu}}P^{\hat{\nu}}=-W(z)(P^0)^2
+W(z)(P^i)^2+(P^z)^2.
\label{restmass}
\end{equation}
Therefore, the momentum $P^z$ of the particle in the direction $z$ of 
the extra space is given by
\begin{equation}
P^z=\sqrt{E^2W^{-1}-M^2-{\bf p}^2W^{-1}},
\label{momz}
\end{equation}
where ${\bf p}=(p^i)$.  The particle is trapped within the 
$(D-1)$-dimensional spacetime, i.e., the motion of the particle along 
the $z$-direction is restricted inside of a finite interval around $z=0$,  
if the energy $E$ of the particle is bounded above by the potential due 
to gravity:
\begin{equation}
E<\sqrt{M^2W(z)+{\bf p}^2}.
\label{ebound}
\end{equation}

One can bring the spacetime metric in Eq. (\ref{dmnwllsol}) for the dilatonic 
domain wall to the form (\ref{warpmet}) by redefining the transverse 
coordinate in the following way:
\begin{eqnarray}
z&=&{\rm sgn}(y){{(D-2)\Delta}\over{2(D-1)+(D-2)\Delta}}Q^{-1}
\left[(1+Q|y|)^{{2(D-1)+(D-2)\Delta}\over{(D-2)\Delta}}-1\right]
\cr
&=&{\rm sgn}(y){{a^2(D-2)^2-4(D-1)}\over{a^2(D-2)^2}}
Q^{-1}\left[(1+Q|y|)^{{a^2(D-2)^2}\over{a^2(D-2)^2-4(D-1)}}-1\right].
\label{newtrnz}
\end{eqnarray}
Then, the dilatonic domain wall metric takes the form (\ref{warpmet}) with 
the warp factor given by
\begin{eqnarray}
W(z)&=&\left(1+{{2(D-1)+(D-2)\Delta}\over{(D-2)\Delta}}Q|z|
\right)^{4\over{2(D-1)+(D-2)\Delta}}
\cr
&=&\left(1+{{a^2(D-2)^2}\over{a^2(D-2)^2-4(D-1)}}Q|z|
\right)^{8\over{a^2(D-2)^2}}.
\label{dwsol2}
\end{eqnarray}
This warp factor for the dilatonic domain wall metric monotonically increases 
as one moves away from the wall and goes to infinity as $|z|\to\infty$, 
if $a^2>4(D-1)/(D-2)^2$, i.e., $\Delta>0$.  (Note, this case includes the 
supersymmetric dilatonic domain walls obtained from the (intersecting) BPS 
branes in string theories through toroidal compactifications, in which 
$\Delta=4/N$.)  So, matter inside of $D$-dimensional spacetime of the 
dilatonic domain wall will always be trapped within the $(D-1)$-dimensional 
sub-spacetime by gravity, even if the extra space with the coordinate $z$ 
can have an infinite extend.  
This mechanism provides with an alternative to the ordinary KK theories where 
the spacetime is assumed to be of the form of the direct product $M_d\times 
K$ of the $d$-dimensional spacetime $M_d$ and some compact space $K$.  On the 
other hand, when $a^2<4(D-1)/(D-2)^2$, i.e., $\Delta<0$, the warp factor 
$W(z)$ decreases at a finite interval $|z|<[4(D-1)/(a^2(D-2)^2)-1]Q^{-1}$ 
around the domain wall, thereby providing with the repulsive potential to 
matter.  (Note, in this case, the spacetime metric (\ref{warpmet}) with the 
warp factor (\ref{dwsol2}) is not well-defined in the region $|z|>
[4(D-1)/(a^2(D-2)^2)-1]Q^{-1}$.)  However, in this case gravity is trapped 
within the domain wall for sufficiently small values of $\Delta$, i.e., 
$\Delta<-2$, as will be discussed in the following.  

This trapping of matter due to gravitational potential well can also be 
seen \cite{go2} by considering the Klein-Gordon equation $\nabla^2\Psi=
M^2\Psi$ for a massive scalar $\Psi$ in the background of dilatonic 
domain wall (\ref{warpmet}) with the warp factor (\ref{dwsol2}).  If we 
let $\Psi=W^{(3-D)/2}(z)\phi(x^{\mu})$, then the above Klein-Gordon equation 
takes the following form of the Klein-Gordon equation for a scalar 
$\phi(x^{\mu})$ in $(D-1)$-dimensional flat spacetime with varying mass as 
one moves along the direction perpendicular to the domain wall:
\begin{equation}
\eta^{\mu\nu}\partial_{\mu}\partial_{\nu}\phi=\left[M^2W(z)+{{D-3}\over 2}
W^{\prime\prime}(z)\right]\phi,
\label{kgeqn}
\end{equation}
where the prime denotes derivative with respect to $z$.  
So, if $a^2>4(D-1)/(D-2)^2$, i.e., $\Delta>0$, then the mass of the 
$(D-1)$-dimensional scalar $\phi(x^{\mu})$ increases as one moves away from 
the domain wall, implying that the scalar $\phi$ is in the attractive 
potential well centered around $z=0$.  When $a^2<4(D-1)/(D-2)^2$, i.e., 
$\Delta<0$, the minimum of the potential well is rather away from the wall, 
i.e., at $|z|=[4(D-1)/(a^2(D-2)^2)-1]Q^{-1}$, which moves away from the domain 
wall as the dilaton coupling parameter $a$ approaches zero.  This implies the 
repulsive potential for the $a^2<4(D-1)/(D-2)^2$ case.  As we will discuss in 
the following paragraph, this critical value $a^2=4(D-1)/(D-2)^2$ for the 
dilaton coupling parameter also corresponds to the critical value below 
[above] which the $\delta$-function potential in the Schr\"odinger equation 
satisfied by the metric  perturbation becomes attractive [repulsive].  
Namely, when $a^2>4(D-1)/(D-2)^2$, i.e., $\Delta>0$, the minimum of the 
potential well for a massive particle or massive scalar is located at the 
domain wall (therefore, matter is trapped due to the attractive potential 
well) and the $\delta$-function potential in the Schr\"odinger equation 
becomes an infinite barrier, but when $a^2<4(D-1)/(D-2)^2$, i.e., $\Delta<0$, 
the minima of the potential well is located away from the domain wall 
(therefore, the potential becomes repulsive) and the $\delta$-function 
potential in the Schr\"odinger equation becomes attractive.

In the following, we study the KK spectrum of graviton in the dilatonic 
domain wall background.  We consider the following general metric describing 
the small fluctuation $h_{\mu\nu}(x^{\mu},u)$ of the $(D-1)$-dimensional 
Minkowski sub-spacetime of the conformally flat $D$-dimensional spacetime 
of the dilatonic domain wall:
\begin{equation}
\hat{g}_{\hat{\mu}\hat{\nu}}dx^{\hat{\mu}}dx^{\hat{\nu}}=
\lambda(u)\left[\left(\eta_{\mu\nu}+h_{\mu\nu}\right)dx^{\mu}dx^{\nu}+
du^2\right],
\label{confflatmet}
\end{equation}
where $|h_{\mu\nu}|\ll 1$.  The conformally flat form of the spacetime 
metric (Eq. (\ref{confflatmet}) with $h_{\mu\nu}=0$) can be achieved from 
the dilatonic domain wall metric in Eq. (\ref{dmnwllsol}) by applying the 
following coordinate transformation:
\begin{equation}
u={\rm sgn}(y){\Delta\over{\Delta+2}}Q^{-1}\left[(1+Q|y|)^{{\Delta+2}
\over\Delta}-1\right],
\label{newtrnr}
\end{equation}
resulting in the conformally flat metric with the following conformal factor:
\begin{eqnarray}
\lambda(u)&=&\left(1+{{\Delta+2}\over\Delta}Q|u|\right)^{4\over
{(D-2)(\Delta+2)}}
\cr
&=&\left(1+{{(D-2)^2a^2-4}\over{(D-2)^2a^2-4(D-1)}}Q|u|\right)^{{8}\over
{(D-2)^2a^2-4}}.
\label{ddcf}
\end{eqnarray}

The $(\mu,\nu)$-component of the Einstein equations (resulting from the action 
which also contains the worldvolume action for the domain wall
\footnote{Note, the scaling symmetry of combined worldvolume and effective 
bulk actions is broken when $\Delta\neq 4$.}) 
in the transverse traceless Lorentz gauge, i.e., $h^{\mu}_{\ \mu}=0=
\partial^{\mu}h_{\mu\nu}$, is approximated to the first order in the 
perturbation $h_{\mu\nu}$ to 
\begin{equation}
\left[\Box_x+\partial^2_u+{{D-2}\over 2}{{\partial_u\lambda}\over{\lambda}}
\partial_u\right]h_{\mu\nu}=0,
\label{eineqttl}
\end{equation} 
where $\Box_x\equiv\eta^{\mu\nu}\partial_{\mu}\partial_{\nu}$.  We let the 
metric fluctuation to be of the form $h_{\mu\nu}(x^{\mu},u)=e^{ip_{\mu}x^{\mu}}
\psi_{\mu\nu}(u)$, where $\eta^{\mu\nu}p_{\mu}p_{\nu}=-m^2$.  Then, Eq. 
(\ref{eineqttl}) reduces to
\begin{equation}
\left[\partial^2_u+{{D-2}\over 2}{{\partial_u\lambda}\over{\lambda}}
\partial_u+m^2\right]\psi_{\mu\nu}(u)=0.
\label{simpeq}
\end{equation}
In terms of new field $\tilde{\psi}_{\mu\nu}(u)=\lambda^{(D-2)/4}(u)
\psi_{\mu\nu}(u)$, Eq. (\ref{simpeq}) takes the following form of the 
Schr\"odinger equation
\begin{equation}
-{{d^2\tilde{\psi}_{\mu\nu}}\over{du^2}}+V(u)\tilde{\psi}_{\mu\nu}=
m^2\tilde{\psi}_{\mu\nu},
\label{scheq}
\end{equation}
with potential
\begin{equation}
V(u)={{D-2}\over{16}}\left[(D-6)\left({{\lambda^{\prime}}\over{\lambda}}
\right)^2+4{{\lambda^{\prime\prime}}\over{\lambda}}\right],
\label{poten}
\end{equation}
where the prime denotes differentiation with respect to $u$.
By substituting the conformal factor (\ref{ddcf}) for the dilatonic domain 
wall metric into this expression, one obtains the following potential 
\begin{equation}
V(u)=-{{1+\Delta}\over{\Delta^2}}{{Q^2}\over{(1+{{\Delta+2}\over{\Delta}}
Q|u|)^2}}+{{2Q}\over\Delta}\delta(u).
\label{potensch}
\end{equation}
This KK potential expression has the similar form as that of the 
non-dilatonic domain wall of the RS model up to the coefficients.  
However, there are some qualitative differences in the KK modes depending 
on the signs of the coefficients in the KK potential (\ref{potensch}).   
In the following, we discuss properties of the KK spectrum for different 
values of $\Delta$ or $a$.
\begin{itemize}

\item $\Delta<-2$ case, i.e., $a^2<4/(D-2)^2$: 
The coefficient in the first term of the KK potential is positive, the 
coefficient of the $\delta$-function term is negative and the coefficient 
in front of $|u|$ in the first term of the KK potential is positive, just 
as in the case of the RS model ($a=0$ case).  So, the KK spectrum consists 
of a single normalizable bound state zero mode (identified as the 
($D-1$)-dimensional graviton) due to the attractive $\delta$-function 
potential and the continuum of massive KK modes which asymptote to plane 
waves as $|u|\to\infty$ (since $V(u)\to 0$ as $|u|\to\infty$) and are 
suppressed as the origin is approached (due to the potential barrier near 
$u=0$).  This implies that the RS model can be extended to the 
$a^2<4/(D-2)^2$ case.   This case cannot be achieved through toroidal 
compactification of intersecting BPS branes in string theories (since 
$\Delta=4/N$ for this case), but can be achieved through the combined 
toroidal and spherical compactifications (when $N(D^{\prime}-p-3)>2$), 
as can be seen from $\Delta$ in Eq. (\ref{adelexp}).  
However, the issues on supersymmetry of such dilatonic domain walls cannot 
be fully answered until one considers corresponding full gauged supergravity 
theory.  

\item $-2<\Delta<-1$ case, i.e., $4/(D-2)^2<a^2<2D/(D-2)^2$:  
The signs of the coefficients are the same as the above case except 
the coefficient in front of $|u|$, which now becomes negative.  So, 
the first term in the KK potential (\ref{potensch}) has the positive 
minimum at $|z|=0$ and increases monotonically, approaching infinity 
as $|u|=-Q^{-1}\Delta/(\Delta+2)$ is reached.  In this case, one rather 
has discrete bound states of massive modes with mass gap.  

\item $-1<\Delta<0$ case, i.e., $2D/(D-2)^2<a^2<4(D-1)/(D-2)^2$: 
The coefficient of the first term in the KK potential becomes negative and 
the coefficient in front of $|u|$ becomes negative, whereas the 
coefficient in the $\delta$-function term still remains negative.  
So, the first term in the KK potential has negative maximum at $|z|=0$ 
and decreases monotonically, approaching negative infinity as $|u|=
-Q^{-1}\Delta/(\Delta+2)$ is reached.  In addition to the continuum of 
massive modes, now there are therefore the continuum of tachyonic modes, 
implying the instability.

\item $\Delta>0$ case, i.e., $a^2>4(D-1)/(D-2)^2$: 
This case includes the case of $\Delta=4/N$, i.e., dilatonic domain walls 
obtained from the (intersecting) BPS branes in string theories through 
toroidal compactifications.  The coefficient in the first term is negative, 
the coefficient in front of $|u|$ is positive and the coefficient in the 
$\delta$-function term becomes now positive.  The first term in the KK 
potential has the negative minimum at $z=0$ and increases monotonically, 
approaching zero as $|z|\to\infty$.  So, in this case, now we have discrete 
bound states of tachyonic modes, in addition to the continuum of massive modes 
without mass gap.  The repulsive $\delta$-function potential at the origin 
provides with infinite potential barrier at $u=0$.  As mentioned previously, 
this case also corresponds to the case in which matter is trapped within the 
$(D-1)$-dimensional sub-spacetime by the attractive potential well due to 
the gravity.  So, dilatonic domain walls obtainable from the (intersecting) 
BPS branes in string theories through toroidal compactification just traps 
matter inside the domain wall without providing with the RS-type scenario.

\end{itemize}

\section{Extreme Dilatonic Branes inside of Extreme Dilatonic Domain Walls}

In this section, we construct fully localized solutions describing extreme 
dilatonic branes within the worldvolume of extreme dilatonic domain walls.  
Schematically, the configurations under consideration are given by the 
following table:
\begin{center}
\begin{tabular}{|l||c|c|c|c|} \hline
{} \ & \ $t$ \ & \ ${\bf w}$ \ & \ ${\bf x}$ \ & \ $y$ 
\\ \hline\hline
brane \ & \ $\bullet$ \ & \ $\bullet$ \ & \ {} \ & \ {}  
\\ \hline
domain wall \ & \ $\bullet$ \ & \ $\bullet$ \ & \ $\bullet$ \ & \ {}  
\\ \hline
\end{tabular}
\end{center}
Here, $t$ is the time coordinate, ${\bf w}=(w_1,...,w_p)$ is the 
overall longitudinal coordinates, ${\bf x}=(x_1,...,x_{D-p-2})$ is 
the relative transverse coordinates for the brane and $y$ is the 
overall transeverse coordinate.  These notations for the coordinates 
will be followed in this and the next sections. 

Such brane configurations are described by the theory which contains the 
$D$-dimensional graviton $g^E_{\mu\nu}$ ($\mu,\nu=0,1,...,D-1$), the dilaton 
$\phi$, $(p+1)$-form potential $A_{p+1}$ with the field strength $F_{p+2}=
dA_{p+1}$, which $p$-brane ($p<D-2$) couples to, and the $(D-1)$-form 
potential $A_{D-1}$, whose field strength $F_D=dA_{D-1}$ dualizes to the 
cosmological constant term which supports domain wall solution.  The 
Einstein-frame action which describes such theory has the following form:
\begin{equation}
S_E={1\over{2\kappa^2_D}}\int d^Dx\sqrt{-g^E}\left[{\cal R}_{g^E}-{4\over{D-2}}
(\partial\phi)^2-{1\over{2\cdot (p+2)!}}e^{2a_p\phi}F^2_{p+2}
-{1\over{2\cdot D!}}e^{2a\phi}F^2_D\right].
\label{einpbrnindw}
\end{equation}

The extreme solution to the field equations of this action describing 
dilatonic $p$-brane localized inside of the dilatonic domain wall has 
the following form:
\begin{eqnarray}
ds^2_E&=&H^{4\over{(D-2)\Delta}}\left[H^{-{{4(D-p-3)}\over{(D-2)
\Delta_p}}}_p\left(-dt^2+dw^2_1+\cdots+dw^2_p\right)\right.
\cr
& &\left.+H^{{4(p+1)}\over{(D-2)\Delta_p}}_p\left(dx^2_1+\cdots+dx^2_{D-p-2}
\right)\right]+H^{{4(D-1)}\over{(D-2)\Delta}}H^{{4(p+1)}\over
{(D-2)\Delta_p}}_pdy^2,
\cr
e^{2\phi}&=&H^{{(D-2)a}\over{\Delta}}H^{{(D-2)a_p}\over{\Delta_p}}_p, 
\cr
A_{tw_1...w_p}&=&1-H^{-1}_p,\ \ \ \ \ \ 
A_{tw_1...w_px_1...x_{D-p-2}}=1-H^{-1}.
\label{pbrdwsol}
\end{eqnarray}
Note, the consistency of equations of motion requires \cite{aeh} that the 
dilaton coupling parameter $a_p$ for the dilatonic $p$-brane is constrained 
to take the following value:
\begin{equation}
a_p={{p+1}\over{(D-2)^2}}{4\over a}.
\label{bpara}
\end{equation}
The harmonic functions $H_p$ and $H$ for the $p$-brane and the domain wall 
satisfy the following coupled partial differential equations:
\begin{equation}
\partial^2_yH_p+H\partial^2_{\bf x}H_p=0,\ \ \ \ \ \ \ 
\partial^2_yH=0.
\label{cpdiffeq}
\end{equation}
A general solution for the harmonic function $H$ to the second equation in 
Eq. (\ref{cpdiffeq}) has the form $H=1+Q|y|$.  However, by applying the 
coordinate translation $y\to y-{\rm sgn}(y)Q^{-1}$ along the $y$-direction, 
one can remove the constant term in $H$.  In this case, the harmonic functions 
that solve Eq. (\ref{cpdiffeq}) are given by \cite{youm}
\begin{equation}
H_p=1+{{Q_p}\over{\left[|{\bf x}|^2+{{4Q}\over 9}|y|^3\right]^{{3(D-p)-8}
\over 6}}},\ \ \ \ \ \ \ \ \  H=Q|y|.
\label{harmfncs}
\end{equation}
Note, these expressions for harmonic functions are valid for any 
values of ${\bf x}$ and $y$.  To bring the harmonic function $H$ to 
the original form, one translates back $y\to y+{\rm sgn}(y)Q^{-1}$ along the 
$y$-direction, leading to the following expressions for the harmonic 
functions:
\begin{equation}
H_p=1+{{Q_p}\over{\left[|{\bf x}|^2+{4\over{9Q^2}}(1+Q|y|)^3\right]^{{3(D-p)-8}
\over 6}}},\ \ \ \ \ \ \ \ \  H=1+Q|y|.
\label{harmfncs2}
\end{equation}

Next, the solution to the field equations of the action (\ref{cosdomanact}) 
describing the pp-wave propagating inside of the extreme dilatonic domain 
wall has the following form:
\begin{eqnarray}
ds^2_E&=&H^{4\over{(D-2)\Delta}}\left[-dt^2+dw^2+(H_{\rm pp}-1)(dt-dw)^2
\right.
\cr
& &\left.\ \ \ \ \ \ \ \ \  +dx^2_1+\cdots+dx^2_{D-3}\right]
+H^{{4(D-1)}\over{(D-2)\Delta}}dy^2,
\cr
e^{2\phi}&=&H^{{(D-2)a}\over{\Delta}},
\label{ppinddw}
\end{eqnarray}
where the harmonic functions $H$ and $H_{\rm pp}$ for the domain wall and the 
pp-wave satisfy the following coupled partial differential equations:
\begin{equation}
\partial^2_yH_{\rm pp}+H\partial^2_{\bf x}H_{\rm pp}=0,\ \ \ \ \ \ \ 
\partial^2_yH=0,
\label{ppdwcpde}
\end{equation}
where ${\bf x}=(x_1,...,x_{D-3})$, and therefore are given by
\begin{equation}
H_{\rm pp}={{Q_{\rm pp}}\over{\left[|{\bf x}|^2+{4\over{9Q^2}}(1+Q|y|)^3
\right]^{{3D-11}\over 6}}},\ \ \ \ \ 
H=1+Q|y|.
\label{ppdwhf}
\end{equation}

Note, we have seen in the previous section that matter is trapped within 
the dilatonic domain wall when $\Delta>0$ and the RS model can be extended 
to the $\Delta<-2$ case.  So, fully localized solutions, presented in 
this section, describes an extreme brane inside of a dilatonic domain wall 
which traps matter [which generalizes the RS model] when the dilaton 
coupling parameter $a$ is such that $\Delta>0$ [$\Delta<-2$].   The former 
case includes the supersymmetric dilatonic domain walls with $\Delta=4/N$, 
which are obtained from the (intersecting) BPS branes in string theories 
through toroidal compactifications.

\section{BPS Solitons inside of BPS Domain Walls in String Theories}

In this section, we construct fully localized supergravity solutions 
describing the BPS solitons inside of the BPS domain walls which are 
obtained by compactifying intersecting branes in string theories.  We shall 
compactify various partially localized intersecting BPS brane solutions 
constructed in Ref. \cite{youm} to obtain such supergravity solutions.  
Because the configurations considered in this section (and also those 
considered in the previous section) describe branes inside of another brane 
and the dimensionality of the overall transverse space is sufficiently small, 
the corresponding supergravity solutions are completely localized.  Moreover, 
these solutions describe the field configurations in the region at arbitrary 
distance from the domain walls.

Generally, the Einstein-frame action for a brane in string theory has 
the following form:
\begin{equation}
S_E={1\over{2\kappa^2_{10}}}\int d^{10}x\sqrt{-G^E}\left[{\cal R}_{G^E}
-{1\over 2}(\partial\Phi)^2-{1\over{2\cdot(p+2)!}}e^{2b\Phi}F^2_{p+2}\right],
\label{einbranact}
\end{equation}
where $G^E_{MN}$ ($M,N=0,1,...,9$) is the Einstein-frame spacetime 
metric, $\Phi$ is the ten-dimensional dilaton and $F_{p+2}$ is the 
field strength of the $(p+1)$-form potential $A_{p+1}$.  Here, the value of 
the dilaton coupling parameter $b$ is -1/2 for the NS-NS 2-form potential 
and $-(p-3)/4$ for the RR $(p+1)$-form potential.  The following string-frame 
action is obtained through the Weyl rescaling transformation $G^E_{MN}=
e^{-\Phi/2}G^s_{MN}$:
\begin{equation}
S_s={1\over{2\kappa^2_{10}}}\int d^{10}x\sqrt{-G^s}\left[e^{-2\Phi}
\left\{{\cal R}_{G^s}+4(\partial\Phi)^2\right\}-{1\over{2\cdot(p+2)!}}
F^2_{p+2}\right],
\label{strbranactrr}
\end{equation}
for the RR $(p+1)$-form potential case, and 
\begin{equation}
S_s={1\over{2\kappa^2_{10}}}\int d^{10}x\sqrt{-G^s}e^{-2\Phi}\left[
{\cal R}_{G^s}+4(\partial\Phi)^2-{1\over{2\cdot 3!}}F^2_3\right],
\label{strbranactns}
\end{equation}
for the NS-NS 2-form potential case.  

We shall compactify some of the longitudinal and the transverse directions 
of the intersecting brane configurations to obtain the configurations 
describing the BPS solitons inside of domain walls.  Since, generally, the 
spacetime metric of intersecting brane solutions are diagonal, the KK Ansatz 
for the string-frame metric is 
\begin{equation}
\left(G^s_{MN}\right)=\left(\matrix{g^s_{\mu\nu}&0\cr 0&g_{mn}}\right),
\label{kkstrmet}
\end{equation}
where $g^s_{\mu\nu}$ ($\mu,\nu=0,1,...,D-1$) is the string-frame metric 
in $D$ spacetime dimensions and $g_{mn}$ ($m,n=1,...,10-D$) is the internal 
metric.  The $D$-dimensional dilaton $\phi$ is expressed in terms of 
the ten-dimensional dilaton $\Phi$ and the internal metric $g_{mn}$ as follows:
\begin{equation}
\phi=\Phi-{1\over 4}\ln\,{\rm det}\,g_{mn}.
\label{ddimdil}
\end{equation}
After obtaining the string-frame metric $g^s_{\mu\nu}$ for a $D$-dimensional 
solution by using the simple compactification Ansatz (\ref{kkstrmet}), one can 
apply the Weyl-scaling transformation $g^E_{\mu\nu}=e^{-{4\over{D-2}}\phi}
g^s_{\mu\nu}$ to obtain the Einstein-frame metric $g^E_{\mu\nu}$ for the 
solution.   In addition to the metric, the dilaton and the form-potential, 
the compactified $D$-dimensional action has additional scalars associated 
with the internal metric $g_{mn}$, unlike the case of dilatonic 
$p$-branes inside of dilatonic domain walls considered in the 
previous section.  As a consequence, the dilaton coupling parameter for 
$p$-branes in string theories are not restricted by the constraint in 
Eq. (\ref{bpara})

Supergravity solutions for branes in string theories have the following 
convenient properties, which simplify the study of brane solutions in $D<10$ 
significantly.  First, the Einstein-frame solution for the spacetime 
metric depends only on the spacetime dimensions $D$ and the dimensionality 
$p$ of the brane, independently of the type of form potential under which 
brane is charged.  Second, in the case of the RR-charged $p$-branes in 
$D<10$, expressions for the spacetime metric and the $D$-dimensional dilaton 
are insensitive to their origin in $D=10$.  Namely, for example, the spacetime 
metric and the $D$-dimensional dilaton for RR 2-brane in $D=6$ have the same 
form whether they are obtained by compactifying one longitudinal direction 
and three transverse directions of D3-brane or they are obtained by 
compactifying four transverse directions of D2-brane.

The universal forms of the Einstein-frame metrics for a $p$-brane and a 
domain wall in string theories in $D<10$ are as follows:
\begin{itemize}
\item $p$-brane:
\begin{equation}
ds^2_E=H^{-{{D-p-3}\over{D-2}}}_p\left[-dt^2+dx^2_1+
\cdots+dx^2_p\right]+H^{{p+1}\over{D-2}}_p\left[dy^2_1+
\cdots+dy^2_{D-p-1}\right],
\label{einbrnsolp}
\end{equation}
where $H_p=1+{{Q_p}\over{|{\bf y}|^{D-p-3}}}$.  This is the $\Delta_p=4$ case 
of the general dilatonic $p$-brane metric in Eq. (\ref{einpbrnsol}).
\item domain wall ($p=D-2$ case of the above):
\begin{equation}
ds^2_E=H^{1\over{D-2}}\left[-dt^2+dx^2_1+\cdots+dx^2_{D-2}\right]+
H^{{D-1}\over{D-2}}dy^2.
\label{eindw}
\end{equation}
This is the $\Delta=4$ case of the spacetime metric for a general 
dilatonic domain wall solution in Eq. (\ref{dmnwllsol}).  
\end{itemize}
Therefore, the Einstein-frame spacetime metrics for the $D$-dimensional 
configurations describing the BPS branes inside of the worldvolume of the 
BPS domain walls in string theories have the following universal forms:
\begin{itemize}
\item  The BPS $p$-brane inside of the BPS domain wall worldvolume:\\
\begin{eqnarray}
ds^2_E&=&H^{1\over{D-2}}\left[H^{-{{D-p-3}\over{D-2}}}_p\left(-dt^2+dw^2_1
+\cdots+dw^2_p\right)\right.
\cr
& &\left.+H^{{p+1}\over{D-2}}_p\left(dx^2_1+\cdots+dx^2_{D-p-2}
\right)\right]+H^{{D-1}\over{D-2}}H^{{p+1}\over{D-2}}_pdy^2.
\label{einstrpindw}
\end{eqnarray}
The harmonic functions $H$ and $H_p$ for the domain wall and the 
$p$-brane satisfy the coupled partial differential equations 
(\ref{cpdiffeq}) and therefore are given by Eq. (\ref{harmfncs}) 
or Eq. (\ref{harmfncs2}).

\item The pp-wave propagating inside of the BPS domain wall:\\
\begin{eqnarray}
ds^2_E&=&H^{1\over{D-2}}\left[-dt^2+dw^2+\left(H_{\rm pp}-1\right)(dt-dw)^2
\right.
\cr
& &\left.\ \ \ \   +
dx^2_1+\cdots+dx^2_{D-3}\right]+H^{{D-1}\over{D-2}}dy^2.
\label{ppindw}
\end{eqnarray}
The harmonic functions $H$ and $H_{\rm pp}$ for the domain wall and the 
pp-wave satisfy the coupled partial differential equations (\ref{ppdwcpde}) 
and therefore are given by Eq. (\ref{ppdwhf}).  
\end{itemize}

The expressions for $D$-dimensional dilaton $\phi$ for various cases are 
as follows:
\begin{itemize}

\item RR $p$-branes (dimensionally reduced from D-branes):  
$e^{2\phi}=H^{(D-2p-4)/4}_p$.

\item magnetic NS $p$-branes with $p=0,1,...,5$ (dimensionally reduced from 
NS5-brane in $D=10$):  $e^{2\phi}=H^{(D-p-3)/2}_p$.

\item electric NS $p$-branes with $p=0,1$ (dimensionally reduced from 
the fundamental string in $D=10$):  $e^{2\phi}=H^{-(p+1)/2}_p$.

\item domain walls compactified from D-branes:  $e^{2\phi}=H^{-D/4}$.

\item domain walls compactified from NS5-brane:  $e^{2\phi}=H^{-1/2}$.  

\end{itemize}

The form potential for a $p$-brane has the standard form $A_{tx_1...x_p}=
1-H^{-1}_p$.  The scalars originated from the internal components of 
the spacetime metric are different for different types of branes and 
compactifications. 

Whereas the Einstein-frame spacetime metrics have the universal 
forms (as discussed in the above), the expressions for string-frame 
spacetime metrics are different for different types of form-fields under 
which the branes are charged.  So, in the following subsections, we present 
the expressions for the string-frame spacetime metric, as well as the dilaton, 
for various cases.

\subsection{BPS solitons within the RR domain walls}

In this subsection, we write down the expressions for the string-frame metrics 
and the dilatons of supergravity solutions describing various BPS solitons 
in the worldvolume of the domain wall charged under the RR form-potential.  

First, we consider the configurations where the RR charged $p$-brane is 
inside of the worldvolume of the RR domain walls.  Supergravity 
solutions for such configurations are obtained by compactifying the 
partially localized supergravity solutions describing two D$p$-branes 
self-intersecting over $(p-2)$-dimensions, D$p$-brane ending 
on D$(p+2)$-brane and D$p$-brane inside of D$(p+4)$-brane.  
The string-frame metric and the dilaton field have the following forms:
\begin{eqnarray}
ds^2_s&=&H^{-{1\over 2}}\left[H^{-{1\over 2}}_p\left(-dt^2+dw^2_1+\cdots
+dw^2_p\right)+H^{1\over 2}_p\left(dx^2_1+\cdots+dx^2_{D-p-2}\right)\right]
\cr
& &\ \ \ \ \ \ \ \ \ \ +H^{1\over 2}H^{1\over 2}_pdy^2,
\cr
e^{2\phi}&=&H^{-{D\over 4}}H^{{D-2p-4}\over 4}_p,
\label{rrpbrninrdw}
\end{eqnarray}
where here and in the following the harmonic functions $H$ and $H_p$ for 
the domain wall and the $p$-brane are given by Eq. (\ref{harmfncs}) or  
Eq. (\ref{harmfncs2}).

Second, the supergravity solutions describing the electric NS-NS charged 
$p$-brane ($p=0,1$) inside of the worldvolume of the RR domain wall 
are obtained by compactifying the partially localized supergravity 
solution describing the fundamental string ending on D-branes.  The 
string-frame metric and the dilaton have the following forms:
\begin{eqnarray}
ds^2_s&=&H^{-{1\over 2}}\left[H^{-1}_p\left(-dt^2+dw^2_1+\cdots+dw^2_p\right)
+dx^2_1+\cdots+dx^2_{D-p-2}\right]+H^{1\over 2}dy^2,
\cr
e^{2\phi}&=&H^{-{D\over 4}}H^{-{{p+1}\over 2}}_p.
\label{nspbrninrdw}
\end{eqnarray}

Third, the  supergravity solutions describing the magnetic NS-NS charged 
$p$-brane ($p=0,1,...,5$) inside of the worldvolume of the RR domain wall 
are obtained by compactifying the partially localized supergravity 
solution describing D-branes ending on NS5-brane.  The string-frame metric 
and the dilaton have the following forms:
\begin{eqnarray}
ds^2_s&=&H^{-{1\over 2}}\left[-dt^2+dw^2_1+\cdots+dw^2_p+H_pdx^2\right]+
H^{1\over 2}H_pdy^2,
\cr
e^{2\phi}&=&H^{-{D\over 4}},
\label{mgnspnsdw}
\end{eqnarray}
where $D=p+3$.  

Fourth, the supergravity solution describing the pp-wave propagating 
in the worldvolume of the RR domain wall is obtained by compactifying 
the partially localized supergravity solutions describing the pp-wave 
propagating in the worldvolume of the D-branes.  The string-frame metric 
and the dilaton have the following forms:
\begin{eqnarray}
ds^2_s&=&H^{-{1\over 2}}\left[-dt^2+dw^2+(H_{\rm pp}-1)(dt-dw)^2+dx^2_1
+\cdots+dx^2_{D-3}\right]+H^{1\over 2}dy^2,
\cr
e^{2\phi}&=&H^{-{D\over 4}},
\label{pppbrninrdw}
\end{eqnarray}
where here and in the following the harmonic functions $H$ and $H_{\rm pp}$ 
for the domain wall and the pp-wave are given by Eq. (\ref{ppdwhf}).

\subsection{BPS solitons within the NS-NS domain walls}

In this subsection, we write down the expressions for the string-frame metrics 
and the dilatons of supergravity solutions describing various BPS solitons 
living in the worldvolume of the domain wall magnetically charged under the 
NS-NS form-potential.  

First, the following string-frame metric and the dilaton of the supergravity 
solution describing the RR charged BPS $p$-brane inside of the worldvolume of 
the NS-NS BPS domain wall are obtained by compactifying the partially 
localized supergravity solution describing the D$p$-brane ending on NS5-brane:
\begin{eqnarray}
ds^2_s&=&H^{-{1\over 2}}_p\left(-dt^2+dw^2_1+\ldots+dw^2_p\right)
+H^{1\over 2}_p(dx^2_1+\ldots+dx^2_{D-p-2})+HH^{1\over 2}_pdy^2,
\cr
e^{2\phi}&=&H^{-{1\over 2}}H^{{D-2p-4}\over 4}_p.
\label{rrpinnsdw}
\end{eqnarray}
where $p=0,1,...,5$.  

Second, the following string-frame metric and the dilaton of the supergravity 
solution describing the electric NS-NS charged $p$-brane ($p=0,1$) in the 
worldvolume of the NS-NS domain wall are obtained by compactifying the 
partially localized supergravity solution describing the fundamental string 
inside of the worldvolume of the NS5-brane:
\begin{eqnarray}
ds^2_s&=&H^{-1}_p\left(-dt^2+dw^2_1+\cdots+dw^2_p\right)+dx^2_1+\cdots+
dx^2_{D-p-2}+Hdy^2,
\cr
e^{2\phi}&=&H^{-{1\over 2}}H^{-{{p+1}\over{2}}}_p.
\label{nspbrinnsdw}
\end{eqnarray}

Third, the following string-frame metric and the dilaton of the supergravity 
solution describing the magnetic NS-NS charged $p$-brane ($p=0,1,...,3$) in 
the worldvolume of the NS-NS domain wall are obtained by compactifying 
the partially localized supergravity solution describing two NS5-branes 
intersecting over 3 dimensions:
\begin{eqnarray}
ds^2_s&=&-dt^2+dw^2_1+\cdots+dw^2_p+H_p\left(dx^2_1+\cdots+dx^2_{D-p-2}\right)
+HH_pdy^2,
\cr
e^{2\phi}&=&H^{-{1\over 2}}H^{{D-p-3}\over{2}}_p.
\label{mgnspbrinnsdw}
\end{eqnarray}

Fourth, the following string-frame metric and the dilaton for the supergravity 
solution describing the pp-wave propagating inside of the NS-NS domain wall 
are obtained by compactifying the partially localized supergravity solution 
describing the pp-wave propagating inside of the NS5-brane:
\begin{eqnarray}
ds^2_s&=&-dt^2+dw^2+\left(H_{\rm pp}-1\right)(dt-dw)^2+dx^2_1+\cdots+
dx^2_{D-3}+Hdy^2,
\cr
e^{2\phi}&=&H^{-{1\over 2}}.
\label{ppinnsdw}
\end{eqnarray}

\end{document}